# AmIE: An Ambient Intelligent Environment for Assisted Living

Marwa Kandil, Reem AlBaghdadi, Fatemah AlAttar
Electrical and Computer Engineering Department
American University of Kuwait
Salmiya, Kuwait
Email: {S00030752, S00029415, S00027443}@auk.edu.kw

Issam Damaj
Electrical and Computer Engineering Department
Rafik Hariri University
Mechref, Lebanon
Email: damajiw@rhu.edu.

*Abstract*— In the modern world of technology Internet-of-things (IoT) systems strives to provide an extensive interconnected and automated solutions for almost every life aspect. This paper proposes an IoT context-aware system to present an Ambient Intelligence (AmI) environment; such as an apartment, house, or a building; to assist blind, visually-impaired, and elderly people. The proposed system aims at providing an easy-to-utilize voice-controlled system to locate, navigate and assist users indoors. The main purpose of the system is to provide indoor positioning, assisted navigation, outside weather information, room temperature, people availability, phone calls and emergency evacuation when needed. The system enhances the user's awareness of the surrounding environment by feeding them with relevant information through a wearable device to assist them. In addition, the system is voice-controlled in both English and Arabic languages and the information are displayed as audio messages in both languages. The system design, implementation, and evaluation consider the constraints in common types of premises in Kuwait and in challenges, such as the training needed by the users. This paper presents cost-effective implementation options by the adoption of a Raspberry Pi microcomputer, Bluetooth Low Energy devices and an Android smart watch.

## I. Introduction

As technology advances further with the growth of the Internet of Things and affects every aspect of modern life it is important to focus on the research and development of new technologies that will improve the lives of those with disabilities [1-5]. The aim of this investigation is to develop a strong relationship between a context-aware system and a wearable device that will improve the assisted living of blind and visually-impaired (BVI) people in unfamiliar indoor environments. According to the World Health Organization (WHO) there are about 253 million people with vision impairment and 36 million who are blind [6]. This makes up a large amount of the population globally and their struggles should not be dismissed. The blind and visually-impaired face many problems in their daily lives which makes their lives more difficult and harder to participate in the society. One of the main issues BVI people must face is difficulty of becoming completely independent in an unfamiliar environment and usually a guide dog or another person assistance is needed to reach their destinations and complete their tasks [7].

The purpose of AmIE is to provide an interactive ambient intelligent and context aware indoor environment for BVI people to help assist and navigate them by utilizing multiple sensor nodes located around the building and a voice-controlled application. The motivation behind this paper exist as it explores the future of mobility and different methods of indoor positioning and navigation which is still a growing field compared to outdoor navigation. It also addresses the pressing issue that BVI people face when trying to navigate in an unfamiliar indoor environment.

Ambient intelligence and context aware systems are rapidly advancing to transform daily life by making the surrounding indoor environments more interactive and interconnected. With the growth of such technology it is the perfect time to advance the field of disability services and assisted living [8]. Context awareness is especially important feature for BVI as it enables the system to act automatically and reduces the burden of excessive user involvement and provides intelligent assistance. To create a context-aware environment three level of data collection from the environment must be achieved these are the participation level, network level and the application level. The participation level includes sensors and other hardware that needs to collect the data from the environment and the user if needed. The network level deals with communication and finally there is the application layer which collects all data gathered and then start preprocessing it. This means filtering and classifying all data while removing the unnecessary noise and integrating it from all the sensors together. Afterwards, context detection is needed using the classified data from different sensors, which will help to infer important information from the environment that the user should be informed about [9].

A variety of different solutions to help assist and navigate BVI are presented in the following part. The three main functions that were represented are indoor positioning and navigation, context aware and ambient intelligence systems, and audio assistive systems.

*A. Indoor Positioning and Navigation*

Indoor positioning and navigation is a relatively new topic when compared to outdoor positioning since using GPS satellites is not possible indoors. Multiple technologies were developed to achieve accurate and fast positioning inside buildings to assist people in navigation. Wireless communication protocols can be used to identify the user location from long distances by using Wi-Fi and for short-medium distances using Bluetooth [10]. One way to locate people indoors is by using Wi-Fi access points distributed around the building [11-12]. This approach locates the user inside the building by representing the floor plans as a graph of connected nodes. In addition to path planning algorithm, Dijkstra, for navigation to the desired location. A similar approach uses Bluetooth devices, when compared to Wi-Fi Bluetooth communication consumes less power and costs. The location of the user can be identified using the Received Signal Strength Indictor (RSSI) from Bluetooth devices around the building by calculating distance based on the characteristics of radio waves progression with change in distance [13]. Which means that distance between the device and the user can be approximated, and thus locating the user in relation to these devices. Another approach for indoor navigation depends on obstacle detection for navigation rather than locating the user inside the building. This is achieved by using both ultrasonic sensors for object detection and image processing for object identification [14-16].

*B. Context-Aware and Ambient Intelligent Environments*

The proposed system by Hudec and Smutny [17] is called RUDO, it is an intelligent ambient environment design for houses of BVI people to assist them in their daily life and parenting. The system support recognition of approaching people, motion detection in the room, and temperature control of the surrounding. Such system is described as context-aware as it communicates with the user through Braille and audio messages. Other systems were designed to assist people in unfamiliar environments yet offer them with relevant data in the surrounding. Prudhvi and Bagani present Silicone Eyes (SiliEyes) [18] which uses a wearable silicone glove to help navigate the user while providing a wide range of information based on the user's current location. Multiple sensors were integrated in the gloves such as a 24-bit color sensor, light sensor, temperature sensor and a SONAR to detect the obstacles. The navigation instructions and sensor data are displayed as audio; however, the system input relies on a touch keypad using Brielle technique for the user to enter their desired location.

*C. Audio Assistive Systems*

The simplest form of communication with BVI people is through a voice-controlled system and audio messages to represent the result of the system. Audio assistive systems require speech-to-text (STT) algorithms for the device to take commands and text-to-speech (TTS) algorithms to interact with the user. Most available speech recognition software is accurate with English language, but they are limited with other languages such as Arabic. Even leading Arabic TTS systems such as Acapela voices and Shakr Arabic-TTS have limited word predictions as well as low quality natural voice [19]. An enduring platform that offers speech recognition is the google cloud speech, it uses machine learning powered by google to recognize speech and its accuracy improves with time. Android OS offers speech recognition using google API [20].

Our contributions in this paper are summarized as follows:

1. We proposed an indoor positioning and navigation algorithm using the RSSI from Bluetooth Low Energy devices distributed around the testing area.
2. We developed a system that uses Raspberry pi as the control unit to provide the user with their current location, navigation directions to a pre-set destination, emergency evacuation directions and sensors data.
3. An android application was developed to provide the user with all features alongside phone call and weather forecast. The application was developed for both English and Arabic speakers.
4. The indoor positioning algorithm was simulated in MATLAB and a real-life testing was conducted to evaluate the whole system.

The organization of the paper is as follows: Section II describes AmIE's design, organization and architecture. Section III explains how the system was implemented. The testing and results are discussed in Section IV, evaluation and analysis are covered in Section V. Finally, Section VI concludes the paper and consider future directions.

II. AmIE ORGANIZATION AND ARCHITECTURE

To integrate BVI people in their environment, an AmIE (Ambient intelligent environment) system is developed to facilitate their interactions with the surrounding. This section focuses on representing the system's functions and technologies. In addition, the Architecture and Organization Diagram is represented to explain the functional requirements of the system clearly and show the final design decisions.

The main purpose of AmIE is to provide an ambient intelligent and context aware environment that helps BVI people to navigate in unfamiliar indoor environments, in addition to informing the user relevant information from the surrounding such as current room temperature, people availability in the room, emergency evacuation alert and display their current location. To integrate the BVI users more in the environment some extra features are developed such as phone call feature and the outside weather alert. To provide the BVI users with the fastest and simplest possible system it must be voice controlled. The system is designed for both English and Arabic speakers where specific key words are used to activate each of the system's requests. The response contains audio messages respectively which may consist of either simple directions, the user's location, sensor data or emergency warning. The AmIE is designed to be a centralized system where the control unit receives requests from the user, collects data from the sensors, and responds back according to the user's request. The Raspberry pi 3 is fast and multithreaded control unit and is used alongside Wi-Fi transceivers to communicate with the sensor nodes to cover wide areas and send data at high data rate. Furthermore, the Raspberry pi processes data from the location nodes and provides directions for navigation when necessary to the user through TCP communication protocol. Multiple nodes are distributed around the building to communicate environment related information such as the location of the user, room temperature and people's availability in the room. These sensor nodes consist of microcontrollers, temperature sensors, and wireless transceivers. For indoor positioning, Bluetooth Low Energy (BLE) devices of standard IEEE 802.15.1 are used to identify the location of the user through measuring the RSSI of the Bluetooth signal by the wearable device. Fig.1 shows the Architecture and Organization Diagram of AmIE.

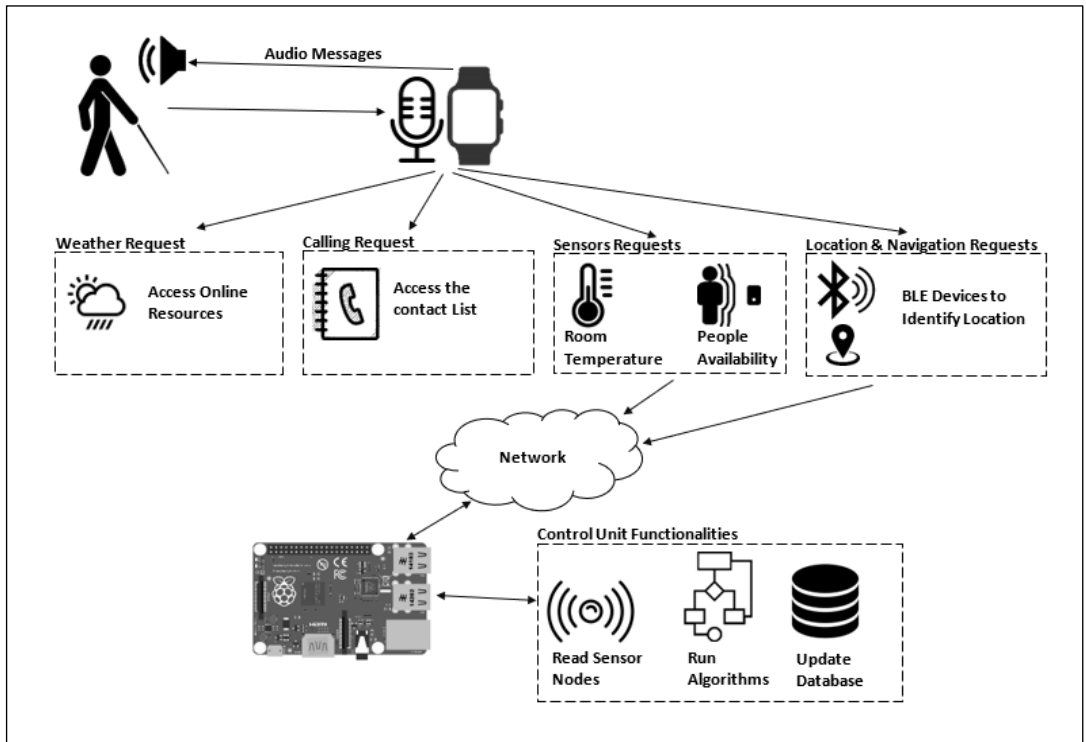

Fig. 1. Architecture and Organization Diagram of AmIE

## III. AmIE Implementation

To demonstrate the feasibility of the proposed system and design a simple prototype was developed and tested. The system implementation is divided into four main parts: the sensor nodes, the location nodes, the control unit and the mobile application. This section discusses the main procedures used to implement each system part and integration of the whole system.

### A. Nodes Implementation

A single sensor node is composed using a sensor module, an Arduino Nano microcontroller, and nrf24l wireless communication module. Two sensors were used, these are analog temperature sensor and HC-SR505 motion sensor. The size of the room defines the number of motion sensors required, however for this prototype four nodes were used to cover up the area and one temperature sensor is needed per room. Fig. 2 shows a single sensor node used in the prototype.

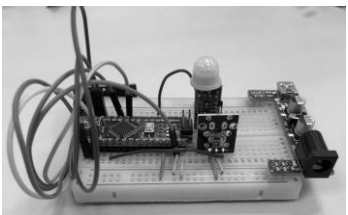
Fig. 2. Circuit of a sensor node

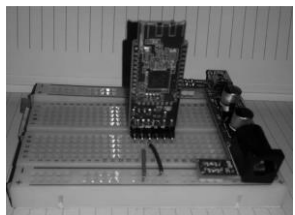
Fig. 3. Circuit of a location node

Location nodes must be placed in a way such that the maximum possible area is covered and provide a range for the RSSI readings below -90dB as otherwise the readings are considered meaningless with distance [13]. Two possible placement strategies are the Point of Interest (POI) and the equidistant strategies. AmIE provides the users with their current location in small areas; the equidistant approach is more suitable since it ensures that there are no gaps between the nodes. In addition, the RSSI readings are guaranteed to be below -90dB which means that the location of the user relevant to the nodes can be identified always. Fig. 3 shows the location nodes containing HM10 CC2541 serial BLE module and a power source.

The testing area is in the American University of Kuwait B building first floor engineering corridor shown in Fig. 4 below. Since the Equidistant approach is the chosen distribution method six nodes are used to cover the area, a set of three nodes would form a triangle to increase the accuracy of the results. The nodes where set on Cartesian coordinates where the first node was located at the origin and the axis were aligned according to its position. The nodes were placed at 2.5 m away in terms of the x-axis, and 5 m apart in terms of the y-axis.

### B. Control Unit Implementation

Raspberry Pi model 3 is defined as a micro-computer with a very fast processor that can perform multithreading which reduces processing time needed during executing the algorithms. In addition, Raspbian operating system is used alongside Python programming language for optimum performance. The Raspberry Pi communicate with the sensor nodes and the mobile application as well as calculating the location of the user and performing navigation. For communication with the sensor nodes, nrf24l wireless transceiver modules are used to cover a wide range. All nodes send to the Pi sequentially through the same channel. The Pi sends messages to the nodes however only one node responds at a time. As for the communication with the mobile application, a TCP connection protocol is established through socket programming. The mobile device sends both requests and array of RSSI values of the user's location to the Pi. The Pi responds based on the request. When the Pi receives requests, it is acting as the server and collects data from sensors. The mobile device is acting as the client in that case placing the request. On the other hand, the mobile device receives data and directions from the Pi.

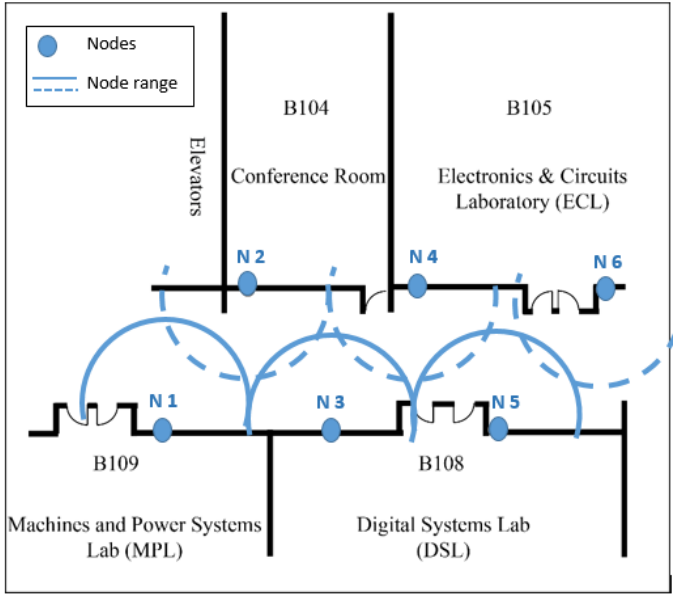

Fig. 4. Map of the testing area with location nodes

### C. Location Algorithm

As mentioned earlier, to identify the location of the user the RSSI is measured from the six Bluetooth Low Energy (BLE) devices to calculate the distance between the user and these nodes to identify where exactly the user is located. The RSS indicates the power level between the BLE device and the user, it is measured in dB and gets more negative as the user moves further from the nodes.

To calculate the distance between the user and the BLE devices multiple reading have been taken from different distances to create an equation that describes the relationship between the RSSI readings and the distance from the BLE devices. Interpolation of the results was done using MATLAB along with the curve fitting function interp1(..) to find the distance equation. The relationship between distance D(x) and RSSI readings x is described in (1).

$$D(x) = -31 * 10^{-5} x^3 - 73 * 10^{-2} x^2 - 1.5 * 10^2 \quad (1)$$

Only the closest three nodes to the user are considered in the calculations to have better prediction of the location and eliminate possible errors. The closest three nodes have the lowest RSSI values and they must be close to each other to reduce error. The distances between the user and these three nodes are calculated using (1) by the control unit. Therefore, the calculated distances are the radii of the circles that represent possible location of the user. The average of the circle intersections is considered the position of the user in terms of Cartesian coordinates and maps it with respect to the nodes. The location algorithm flow chart is shown in Fig. 5.

Navigation is preformed when the user requests a specific destination. The path is predefined since the testing area is simple. The current location of the user is determined, and the path is displayed in simple directions such as forward, turn left etc. The location of the user is constantly measured to check if the user reached the destination.

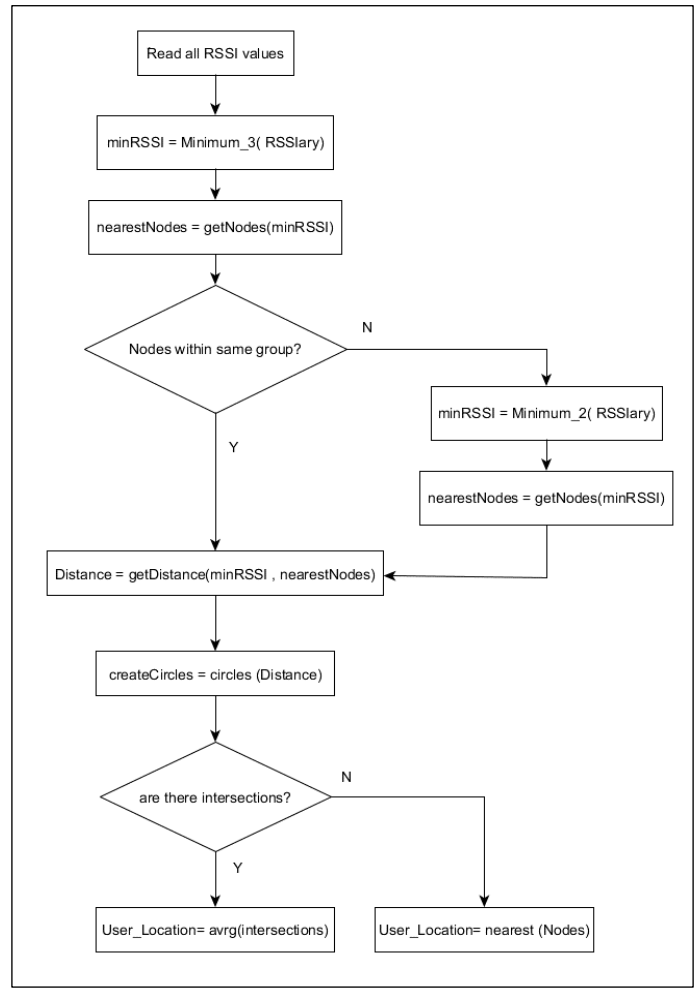

Fig. 5. Location Algorithm flow chart

### D. Mobile Application

The main functions of the Android application are voice recognition, voice responses and measuring the RSSI. Voice recognition and responses can be in both the English and Arabic languages. The voice recognition was implemented through the Google Speech API [21]. However, the voice responses that the user receives after inputting a request are prerecorded messages. The user needs to input specific keywords to get the appropriate responses. Measuring the RSSI values is done through the Android SDK which can get the RSSI from the surrounding BLE modules. The application scans for all six BLE devices, measures their RSSI values accordingly and send the values in an array to the control unit. In addition, the application has a direct access to the contact list such that if the user requests to call someone for assistance that person will be called directly.

## IV. TESTING AND RESULTS

The testing of the system is performed by blind folded participants and the testing area was in the engineering corridor in B building of the American University of Kuwait. Participants were asked to test all features of the system and these include: positioning, navigation, outside weather, calling feature, emergency feature and sensor nodes. Multiple tests were conducted, and the responses of the participants was recorded.

According to the tested blind folded participants the system was user-friendly and easy to use as placing a request was not difficult. Fig.6 shows the user interface featuring the button to place a request. Participants tested the positioning feature in different locations in the corridor and results displayed determined their nearest point of interest (rooms). The location coordinates were displayed on the Raspberry pi screen as shown in Fig. 7. The calculated error shows an average of ±11.6% in the x axis and ±14.7% in the y axis. This means that the system is almost 85% accurate in both axis and the system can identify which nodes are closest to the user thus estimate his/her location. The navigation feature was tested in the corridor from multiple locations. Participants stated that the directions were clear and easy to follow, and the system indicated when they reached their destination. Example of directions is displayed in TABLE 1. Furthermore, the emergency evacuation was tested, and the user was navigated to the exist location marked as node 6 on the map and the original location was displayed on the screen connected to the raspberry Pi as shown in Fig. 8.

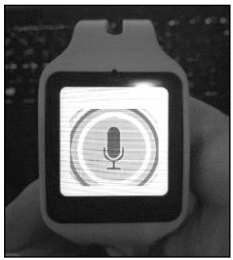

Fig. 6. User Interface of the application

The sensor nodes features were tested in one room in the same floor. According to the participants the response time was quick and clear in both languages. In addition, they said that data of the room temperature and people availability provided them with good idea of their surrounding environment. Furthermore, the calling request works well in calling numbers that are already saved in the contact list. The outside weather functionality was instant; however, some delay was encountered when the sensor battery level dropped. Despite this, participants had to memorize commands needed to activate the requests and their devices had to be constantly connected to the internet to maintain the communication.

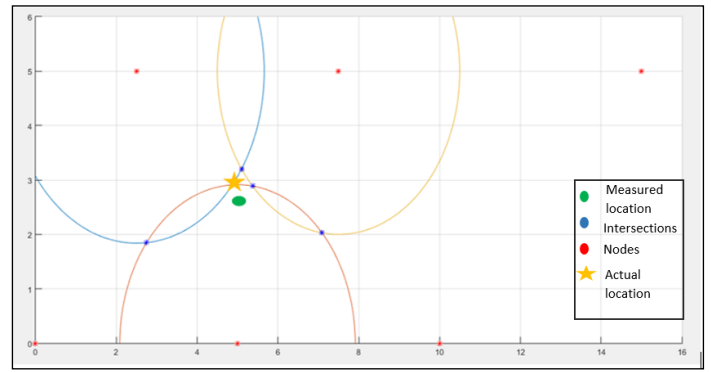

Fig. 7. Result of user's location

TABLE I. RESPONSES TO THE USER

| Request | Responses | |
|---|---|---|
| | *English Response* | *Arabic Response* |
| Navigate Digital Lab | Move Forward<br>Move Forward<br>Move Forward<br>Turn Left<br>You have reached your distination | تحرك الى الامام<br>تحرك الى الامام<br>تحرك الى الامام<br>تحرك الى اليسار<br>لقد وصلت الى المكان المطلوب |

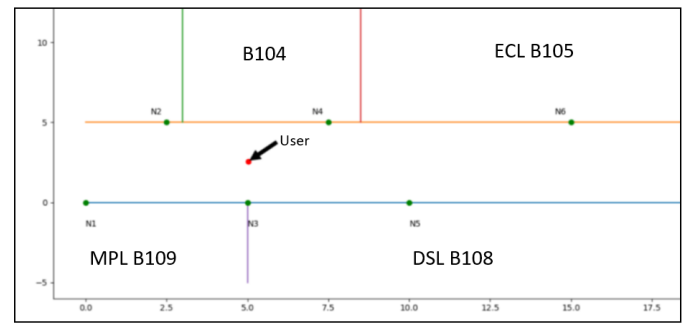

Fig. 8. Raspberry Pi screen output of user's location during emergency evacuation

## V. AmIE EVALUATION

For evaluation of the proposed system a comparison study was conducted against a system proposed by Adam Satan and Zsolt Toth [22]. They presented a Bluetooth-based indoor positioning mobile application, specifically designed for offices to notify the users of the nearby offices. Their positioning algorithm is similar in AmIE. However, AmIE does not take into consideration the height of the Bluetooth beacon as this system does. In addition, AmIE's mathematical model that relates the RSSI values to the user's distance from the nodes is generated through Interpolation using MATLAB, whereas this system uses the Log-distance Path Loss Model. Both systems tested through a hallway making the comparison between both system accuracy fair. system produced an accuracy percentage of 88% which is not far from the 85% accuracy of AmIE especially that the user's exact location can be estimated correctly. This shows that the location algorithm has great potential alongside improving it in the future. In addition, AmIE provides the BVI users with wide range of ambient intelligent functionalities that increases their awareness of the environment.

Any system faces multiple limitations and challenges during the designing and implementation phases, so does AmIE. However, these challenges were overcome by applying appropriate solutions. Transmitting ranges of the sensor nodes were one of the limitations faced, as their range was limited to 10m. A further improvement is to use a wider range antenna to cover wider areas. Another challenge faced is that speech recognition at times can be inaccurate especially in crowded areas, as a solution a set of headphones with embedded microphone was used to reduce the error. One of the main constraints to the system is the failed connection with the internet which caused the whole system to stop working. In addition, the prototype was designed for a rectangular hallway which makes the system limited to the shape of the room.

## VI. CONCLUSION

AmIE is an IoT system designed for the assistance of BVI people to improve their quality of life and ensure that they can reach their destinations in unfamiliar indoor environments in a safe and effective way. The addition of extra features other than the main positioning feature ensure that the system encompasses many elements that will be beneficial to the user. The testing of the system proved that it is reliable and functioning as expected. Future works include testing the system with BVI people so that their experience can be evaluated and used to improve the system. Moreover, the system will be tested in different locations and in areas of different shapes and sizes to understand how the area affects the accuracy of the system.